\documentclass{sig-alternate-05-2015}

\usepackage{balance}
\begin{document}

\newenvironment{definition}[1][Definition]{\begin{trivlist}
\item[\hskip \labelsep {\bfseries #1}]}{\end{trivlist}}

\setcopyright{rightsretained}





%

\title{Portable Ontological Expressions in NoSQL Queries}
%
%
%
%
%

\numberofauthors{2} 
%
\author{
%
%
\alignauthor
Suresh K. Damodaran\\
       \affaddr{MS M335, MITRE}\\
       \affaddr{202 Burlington Rd.}\\
       \affaddr{Bedford, Massachusetts, USA}\\
       \email{sdamodaran@acm.org}
\alignauthor 
Pedro A. Colon-Hernandez \\
       \affaddr{MIT Lincoln Laboratory}\\
       \affaddr{244 Wood St.}\\
       \affaddr{Lexington, Massachusetts, USA}\\
       \email{pedro.colon-hernandez@ll.mit.edu}
}
\date{30 Dec 2015}

\maketitle
\begin{abstract}

A significant barrier to the portability of queries across diverse physical implementations of large data stores, especially NoSQL data stores, is that the queries reference the physical storage attributes, such as the table and column names.
In this paper, we describe a technique for embedding ontological expressions called Address Expressions, or A-Expressions, in NoSQL queries to improve their portability across diverse physical implementations. We discuss an implementation of such queries over a MongoDB data store of the Enron email corpus with examples, and conduct a preliminary performance assessment. 

\end{abstract}

%
%


%
%

%
%



\section{Introduction}

Big Data Analytics applications use NoSQL data stores of logs for processing large amounts of data \cite{Fisher:2012}. Portability of queries across implementations, especially when embedded in algorithms for data processing, is a difficult challenge for NoSQL data stores. One significant cause of this difficulty is that the current query languages and programming interfaces for NoSQL use the physical table, column, and column family names of the data store for specifying the data \cite{mahout:2011,PIG,Thusoo:2009}. Since there are no standardized names for these columns or column families, every project is free to assign any names.  The lack of standardization in naming can cause difficulties in porting an algorithm written and found very useful in one project to another project. Analyzing the portability of an algorithm is also difficult because it requires careful study of both the names and their semantics in both the original data store and the new data store.   In the rest of the paper, we will use the more general term \textit{field} instead of \textit{column}, since a log record consists of a set of fields, whereas  \textit{column} is a term used in the storage of log records in NoSQL data stores. We will also use the term \textit{table} to refer to a collection of records of log data. 

Standardization of the field names is a hard problem because the interpretation of the data can be as varied as the users of the data. Take for example a field that will contain source IP address. In one log this field might be named \textit{src\_ip},  whereas in another it might be named \textit{source\_ipv4} to differentiate it from a field named \textit{source\_ipv6}. \textit{src\_ip} is short, but \textit{source\_ipv4} is more descriptive. Similarly, does \textit{source} mean an entity outside an organization? Often the answers to these questions vary, causing confusion and endless debates on a standard name for the field. We argue that  standardizing a field name is not productive, because it is hard, if not impossible, to name a field considering all possible uses and interpretations of the data in a field. Furthermore, the same data may be interpreted differently in the future in future applications. Therefore, we need to consider another approach. 

We propose an alternative approach for portability of embedded queries. Our approach uses an ontology based expression we term Address Expressions, or, A-Expressions that can be embedded in any NoSQL query languages. We term this enhanced language as Knowledge Query Language (KQL). The ontology layer rewrites KQL queries into  queries containing physical attributes of the data store such as the tables, columns, and rows. We demonstrate our  embedding technique for A-expressions using SQL, with a SQL driver \cite{mason:2005} for MongoDB datastore.

The contributions of this paper include (1) a novel approach for developing an ontology, based on the separation of mutable and immutable attributes of the information content in log files, (2) implementation of a semantic layer for rewriting KQL queries into SQL queries with data store attributes using this ontology, (3) demonstration with examples the rewriting process of KQL queries, and (4) a preliminary performance evaluation of KQL queries.

The rest of the paper is organized as follows. In Section 2, we describe how an ontology can be created based on the mutable and immutable attributes of data for mapping to the physical, implementation specific attributes of data stores. In Section 3, we provide examples of KQL queries, and a summary of the implementation of these queries. Section 4 contains the analysis of performance of KQL queries. Related work is discussed in Section 5, and we conclude the paper in Section 6.

\section{Immutable and Mutable Aspects of Data}	
We show now how the name of a field encodes both the immutable type and mutable context of the data stored in the field with a more detailed example. Consider the field we examined earlier, \textit{src\_ip}. This name encodes the type of the information (here,  \textit{IPAddress}), and its context (here,  \textit{src}). Note that another field name,  \textit{source\_ip}, may also have the same type,  \textit{IPAddress}, but the context is named  as \textit{source} instead of \textit{src} (see Fig. 1).  \textit{src\_ip} may have an additional implicit context of  \textit{ext} or  \textit{int}, for external and internal, respectively. Similarly, for the field  \textit{dst\_ip}, the type is the same as that of  \textit{src\_ip}, which is  \textit{IPAddress}, but has a different context,  \textit{dst}. In another log, the field  \textit{server\_ip} has the type  \textit{IPAddress}, but has a different context,  \textit{server}.  Yet another field may be  \textit{src\_port} that has a type  \textit{PortNumber}, and has the context  \textit{src}.  To use the fields  \textit{src\_ip},  \textit{dst\_ip},  \textit{src\_port}, and  \textit{server\_ip} effectively, the analyst needs to have a detailed understanding of the type and context information for each field that is present in the logs, and the relationships among the fields. Therefore, it would be reasonable to define type and context as separate concepts that can be together used to reference a field.

\begin{figure}
\centering
\includegraphics[width=0.45\textwidth]{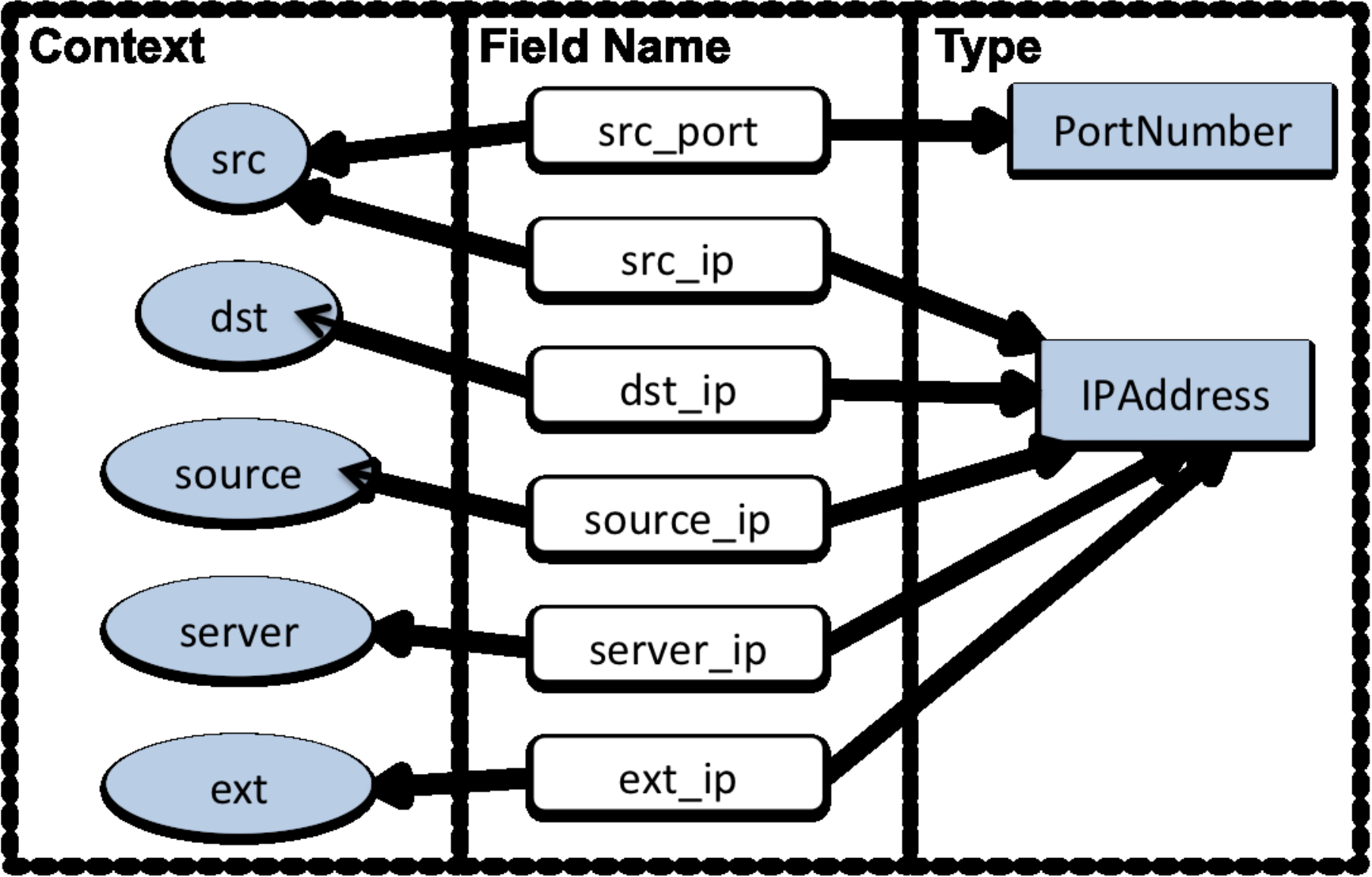}
\caption{Types and Contexts.}
\end{figure}
 
To explicitly denote type as the immutable aspect of the data, we define \textit{Dimension}.  We use the term \textit{Dimension} instead of \textit{type} to avoid any confusion with programming language data types. For example, \textit{PortNumber} can be a Dimension of  \textit{src\_port}, whereas in a programming language, this field likely will have the data type of  \textit{Integer}. The range of values a Dimension may depend on its data type, in addition to other constraints. It is possible that fields are assigned default data types such as \textit{String}. However, we do not consider these system-assigned default data types as Dimensions. 

\begin{definition}
 1. A  \textit{Dimension} is an immutable and inherent attribute of a field that, once assigned, will not change. A Dimension may have an implied, or an enumerated  range of values.
\end{definition}

In contrast to a Dimension, the context of the data stored in a field is mutable in the sense that the context may change after its assignment to a field. The context may be based on the current or potential use of a field, or based on the origin of the information stored in the field. The use of data may change over time, and therefore, it is extremely hard to anticipate all the possible uses of a field. Therefore, it would really help the analyst to be able to assign additional contexts over time, or even add personalized contexts. Presently, we define the term Tag to denote a context because tagging is a widely used collaboration approach in various web applications \cite{Begelman:2006}.  A helpful analogy is to think of Dimension as a noun and a Tag as an adjective. Since both Dimension and Tags are assigned, it is possible that some fields in a data store do not have either. 

\begin{definition}
2. A  set of \textit{Tags} represents the mutable attributes of data.  Tags may be associated or disassociated with data at any time.
\end{definition}

\begin{definition}
3. An \textit{Address Tuple} <D, S> can be used to specify a field, where D is a Dimension, and S is a set of Tags.
\end{definition}

Often data from fields in the same row of a table needs to be retrieved. Such retrieval is accomplished by defining the concept of a \textit{DimensionSet} which consists of a set of Dimensions. If a physical table contains columns with every one of the Dimensions in the DimensionSet, then all the rows of those columns may be retrieved.

\begin{definition}
4. A \textit{DimensionSet} is a non-empty set of Dimensions.
\end{definition}

In the next section we apply the concepts developed in this section to develop an ontology for email messages. 

\subsection{An Ontology for Email Messages}

Email message structures are well understood, and therefore we use Enron email message corpus \cite{Klimt:2004} to demonstrate the use of the concepts described in the previous section. Ontology, according to W3C consortium \cite{W3C}, is "the definition of terms used to describe and represent an area of knowledge."  The terms include names of the concepts, and relationships among themselves. In the email ontology, the terms are instances of Dimension, Tag, DimensionSet, the physical table where the data is stored, and its fields. We store the entire email message corpus in a single table, \textit{email\_message\_table}. This table has the fields shown in Fig. 2.
 
\begin{figure}
\centering
\includegraphics[width=0.30\textwidth]{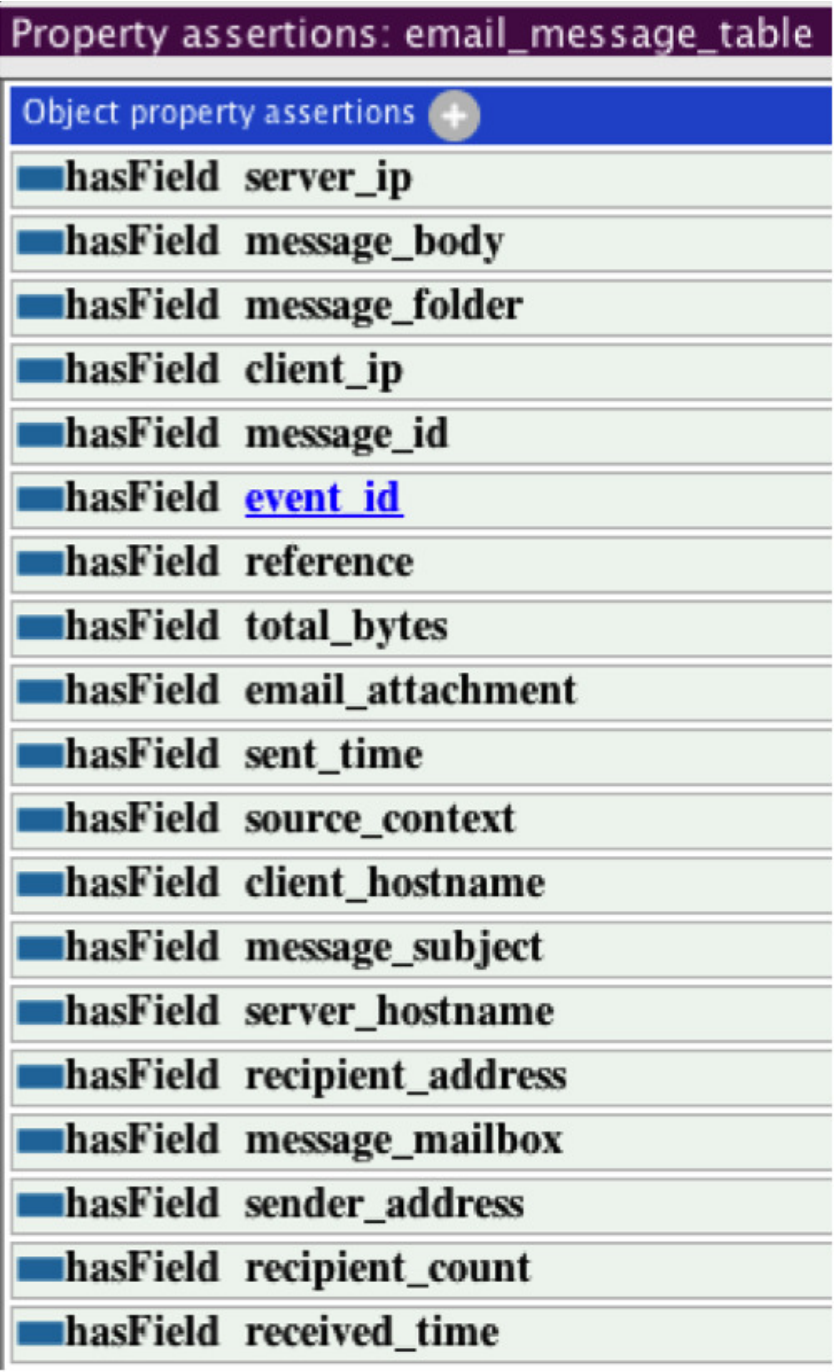}
\caption{Email Message Fields.}
\end{figure}
 
The Dimensions and Tags in the email ontology are shown in Fig. 3. One may observe that \textit{lastName} in Fig. 3 is a Dimension in this ontology. An alternate approach could be to have \textit{name} as the Dimension, and have \textit{last} and \textit{first} as Tags. The definition of a standardized email ontology with Dimensions and Tags will be a community effort. Another observation is that there are 19 Dimensions, and 14 Tags, and these can form 19x14 = 266 Address Tuples to address as many fields, if all Tags were to be meaningfully combined with all Dimensions to form Address Tuples! Of course, in reality, not all those Address Tuples would have a corresponding field, and a field may have multiple Tags. Yet, it is obvious that there are many nuanced ways to identify a field, using Dimensions and Tags, and standardizing the names of 19 Dimensions and 14 Tags may be easier than standardizing, in the worst case, 266 field names.

 \begin{figure}
\centering
\includegraphics[width=0.35\textwidth]{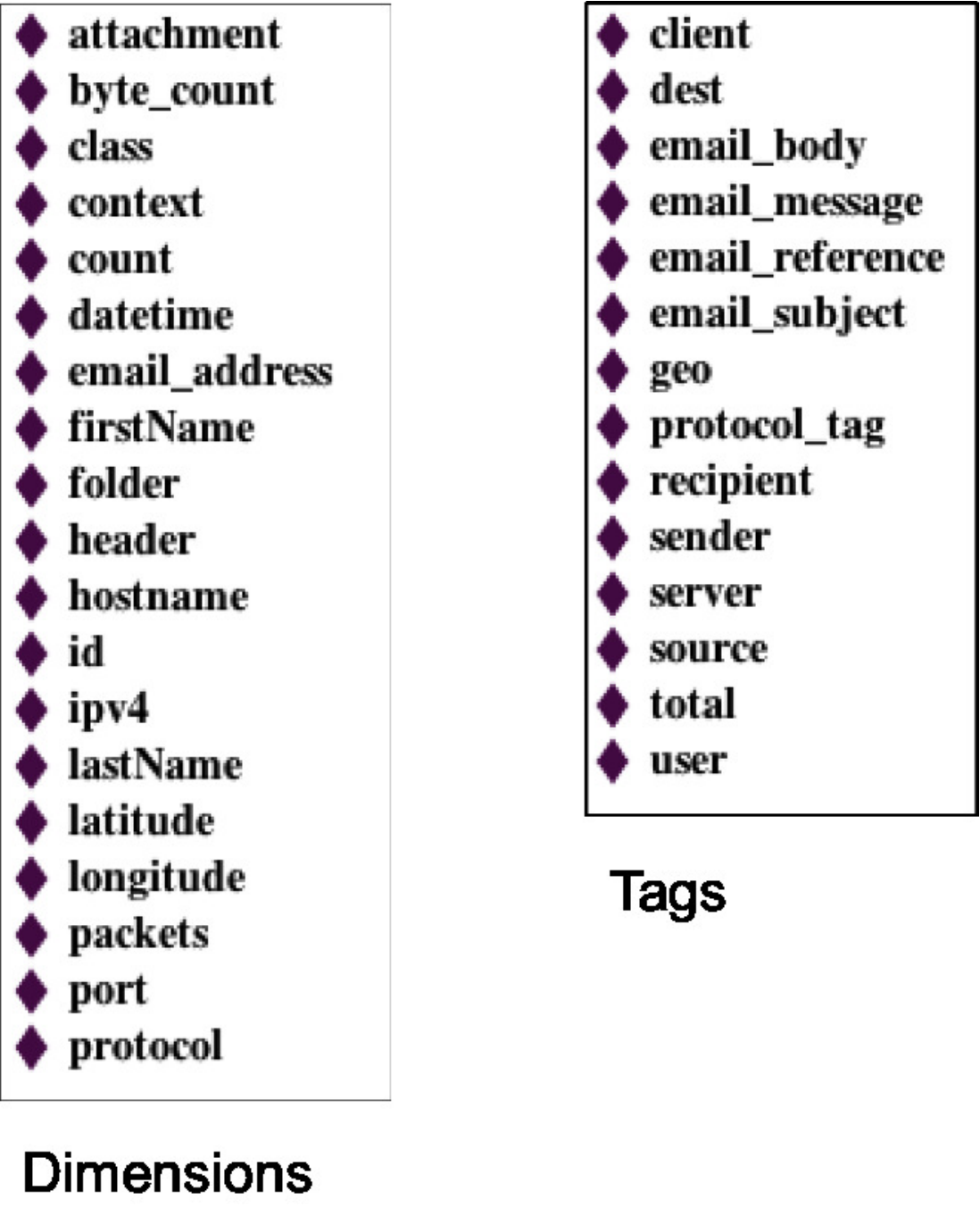}
\caption{Email Dimensions and Tags.}
\end{figure}

The DimensionSets defined for the email ontology are listed in Fig. 4. There are two DimensionSets defined, one named \textit{email\_event}, and another named \textit{emailmessage}. How many DimensionSets are defined, and which Dimensions are included in them, are dependent on current log structure, and their anticipated use.
 
\begin{figure}
\centering
\includegraphics[width=0.40\textwidth]{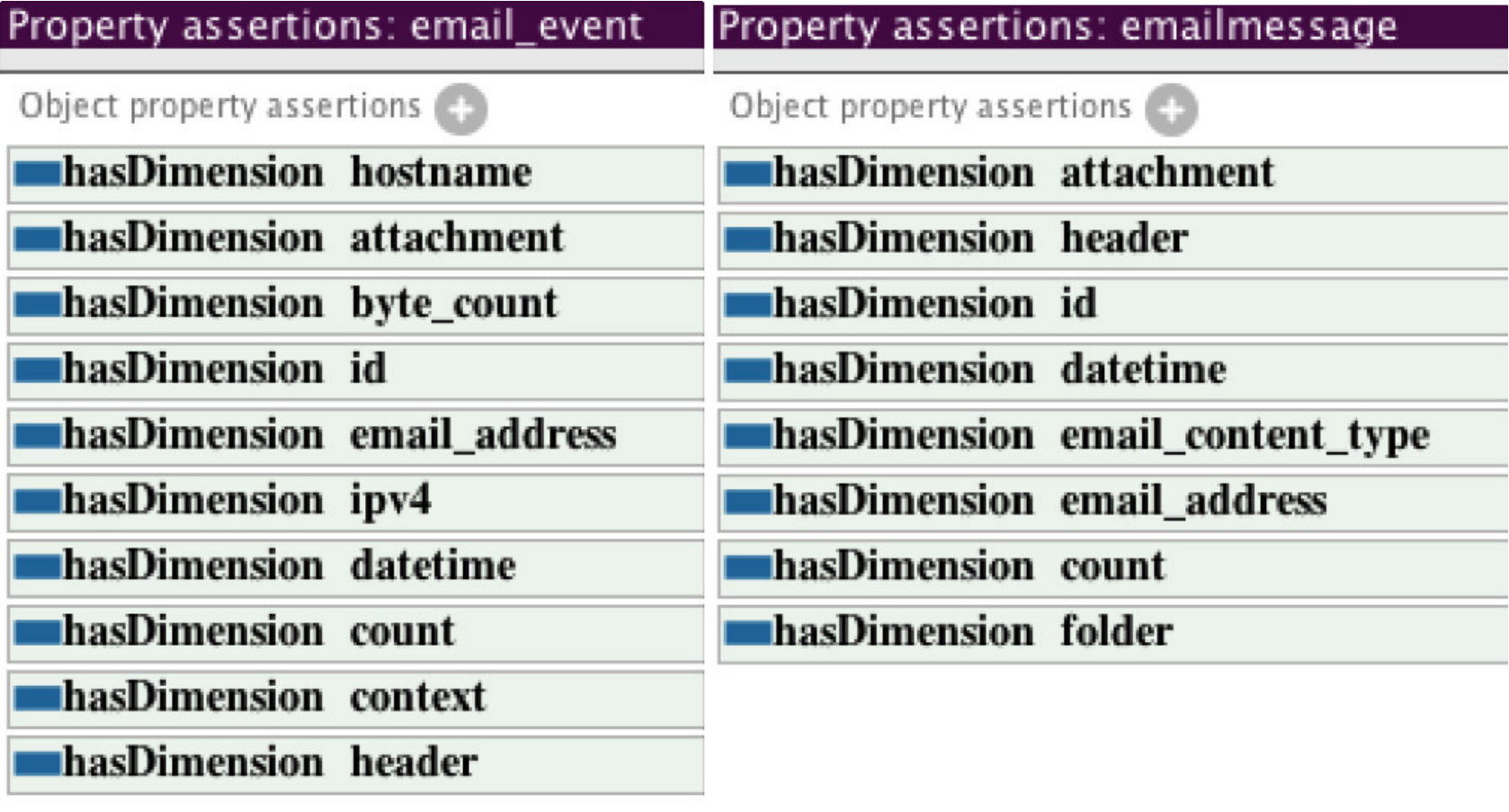}
\caption{Email DimensionSets.}
\end{figure} 

In the next section, we define expressions containing Dimensions, Tags and DimensionSets instead of the physical names of the tables, and show how they are rewritten into SQL queries using an ontological layer. We term such expressions \textit{Address-Expressions}, or \textit{A-Expressions}. Since A-Expressions are based on the knowledge embodied in the ontology,SQL extended with embedded A-Expressions is termed Knowledge Query Language (KQL). We discuss KQL next.

\section{Knowledge Query Language}
We presently examine some KQL queries, how they are rewritten into plain SQL queries, and further into queries specific to a data store, in this case, a MongoDB data store.
The A-Expressions used in the following KQL queries are based on the email ontology described in Section 2. In the A-Expressions below, \textit{emailmessage} is a DimensionSet, and is described in Fig. 4. For the purpose of this discussion, we will explain the needed syntax and semantics of A-Expressions as required. For detailed syntax and semantics of A-Expressions, please see \cite{sureshkql:2016}.

\subsection{SELECT DISTINCT Example}
This query, described below, is for all the email addresses of people who have sent emails. The query has four A-Expressions. The A-Expression \textit{ALL*email\_address*\_:source} specifies all fields with \textit{email\_address} as Dimension and \textit{source} as Tag. For the purpose of this paper we can ignore \mbox{\_:} which is used to specify the default TagScheme \cite{sureshkql:2016}, since we use only a single TagScheme. In the FROM clause, \textit{ALL/emailmessage}, specifies all the data from rows from physical tables that have every one of the Dimensions in the DimensionSet \textit{emailmessage} as columns. The A-Expression in the WHERE clause specifies all email messages in the \textit{sent} or \textit{sent\_items} folders.

\begin{verbatim*}
SELECT DISTINCT ALL*email_address*_:source 
FROM ALL/emailmessage
WHERE ( ALL*folder = 'sent_items' OR 
ALL*folder = 'sent')
\end{verbatim*}

The above KQL statement is rewritten in SQL as follows using the field names shown in Fig. 2:
\begin{verbatim*}
SELECT DISTINCT sender_address  
FROM email_message_table  
WHERE (message_folder = 'sent_items' OR  
message_folder = 'sent' ) 
\end{verbatim*}

A corresponding MongoDB query is as follows: 
\begin{verbatim*}
db.email_message_table.distinct("sender_address", 
{ "$or" : [{ "message_folder" : "sent_items"}, 
{ "message_folder" : "sent"}]})
\end{verbatim*}

The result of this execution is a set of the distinct email addresses from email senders as shown below:
\begin{verbatim*}
mark.haedicke@enron.com
vince.kaminski@enron.com
steven.kean@enron.com
tori.kuykendall@enron.com
...
\end{verbatim*}

\subsection{Nested KQL Query Example}
Let us consider a nested KQL Query example now:
\begin{verbatim*}
SELECT ALL*[emailmessage] 
FROM (SELECT * FROM ALL/emailmessage 
WHERE 
ALL*email_address*_:sender='susan.scott@enron.com') 
as example 
WHERE ( ALL*datetime*_:sender >= 
'2000-01-01 00:00:00-07:00' AND 
ALL*datetime*_:sender < 
'2003-01-01 00:00:00-07:00' )
\end{verbatim*}
In this query, we want to get all the email messages that were sent by  \textit{susan.scott@enron.com} in a specific timeframe.  The A-Expression \textit{ALL*[emailmessage]} specifies all fields that have any of the Dimensions in DimensionSet  \textit{emailmessage}. The \textit{ALL*email\_address*\_:sender} specifies all fields with Dimension \textit{email\_address}, and Tag \textit{sender}. Finally, \textit{ALL*datetime*\_:sender} specifies all fields with Dimension  \textit{datetime}, and Tag  \textit{sender}. The corresponding rewritten SQL query is below.
\begin{verbatim*}
SELECT message_id,sent_time,recipient_address,
message_folder, received_time,message_body,
email_attachment,sender_address,
recipient_count,message_mailbox,message_subject  
FROM (SELECT * FROM  email_message_table  
	WHERE sender_address = 'susan.scott@enron.com')  
	as example  
WHERE (sent_time >= '2000-01-01 00:00:00-07:00' 
AND sent_time < '2003-01-01 00:00:00-07:00')
\end{verbatim*}
For brevity, we do not provide the corresponding MongoDB query or results. It should be noted that A-Expressions cannot be used to specify tables or fields that have no Dimensions, or Tags. In such situations, KQL queries can use \mbox{\textit{SELECT *}} to specify fields of a table, irrespective of whether all the fields of the table have Dimensions or Tags. Also, the keyword \textit{ALL} may be used, which specifies all tables in the system.

\subsection{Implementation of the KQL Layer}

The KQL queries described in the previous section are rewritten into SQL queries using the email ontology. Fig. 5 describes how the KQL queries are implemented. The algorithm we implemented for demonstrating the execution of KQL queries was a community detection algorithm written in Python \cite{Girvan:2002}. The KQL queries were embedded in the Python source code, and since KQL implementation was done using Java, the queries were run using Java based libraries. An SQL parser \cite{psp:2012} was modified to extract and replace embedded A-Expressions. This SQL parser would make sure the SQL components of the KQL queries have valid SQL syntax. The KQL queries are then processed by the KQL layer that includes a KQL translator or rewriter. The rewriting of the A-Expressions, as described in the previous section, is done by the KQL translator using information in a Knowledge Registry. The Knowledge Registry implements the email ontology, using JSON files. The output of the KQL layer is an SQL query that is processed by UnityJDBC driver \cite{lawrence:2000,mason:2005} that can translate the SQL query to a query that can be executed over the data store implemented using MongoDB. The result of the query is then written to a file.
\begin{figure}
\centering
\includegraphics[height=3in, width=2.5in]{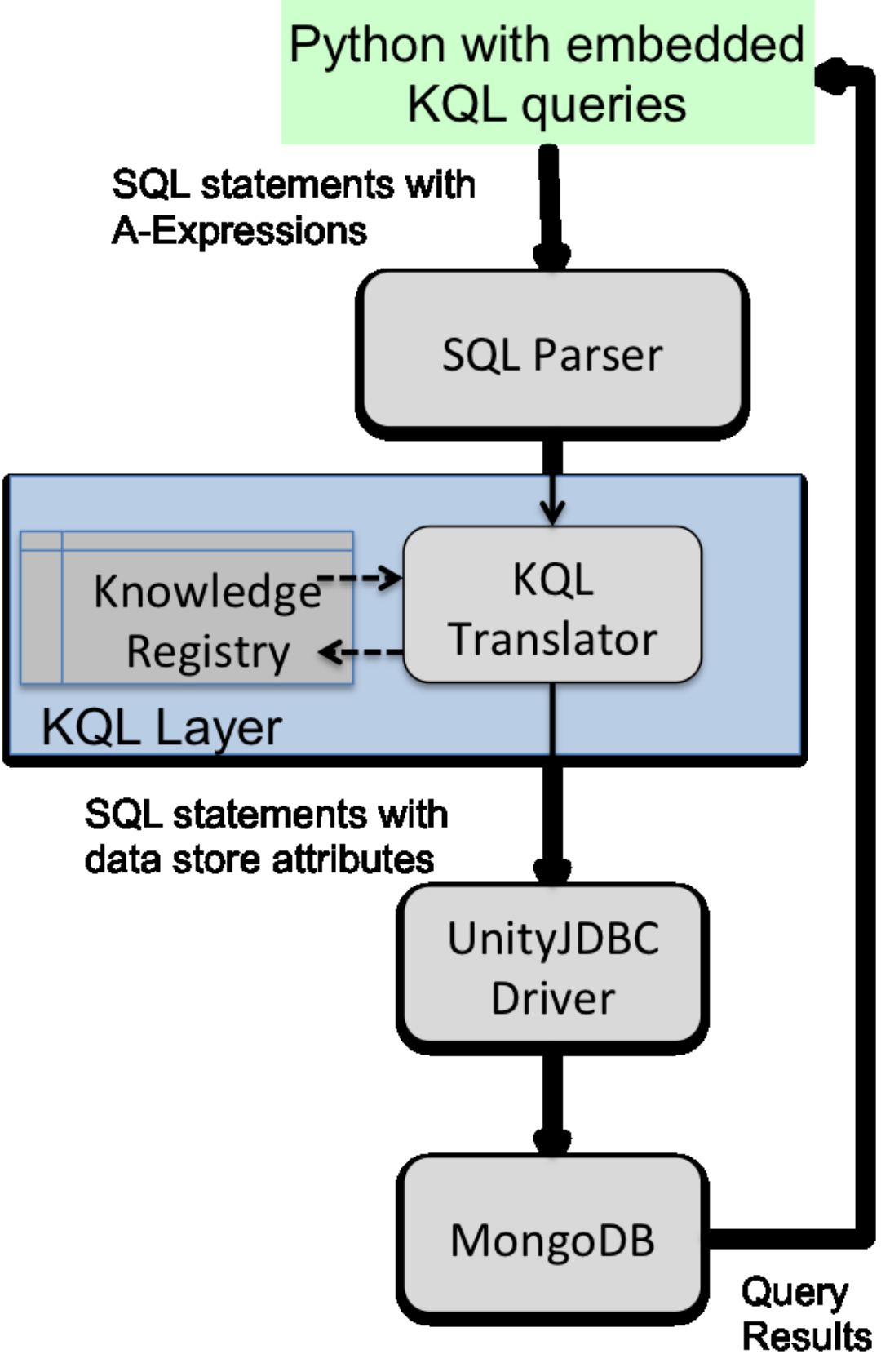}
\caption{KQL Query Implementation.}
\end{figure} 
In the next section we describe the results of a preliminary performance evaluation to assess the impact of the KQL layer on generation of the query results.

\section{Performance Evaluation}
We evaluated the performance impact of the KQL layer by conducting experiments with the same community detection algorithm over subsets of the Enron corpus \cite{Klimt:2004} with and without using the KQL layer. This algorithm has a few parameters that can be set to change its running time. One of them is obviously the size of the data that the algorithm is run on. Another parameter is to define the strength of an email connection through a minimum number of messages sent by a person to a recipient \cite{tyler:2005}. The default value we use for this parameter is 10. The number of initial senders is another adjustable parameter to the community detection algorithm. By default, the number of unique email addresses of senders is used as the total number of initial senders \cite{tyler:2005}. We vary these parameters to conduct our experiments as discussed below.

We use modified versions of the existing scripts \cite{Nehl:2011} to load the Enron corpus data into MongoDB. The modifications are to remove forwarded parts from a message, since in our analysis this information is not needed. We also imported data only from the \textit{sent} and \textit{sent\_items} folders for each user of the corpus because these folders contained the greatest amount of unique messages from the users for whom information is available in the corpus. We also segregated the corpus data in periods of 5,10,15,20,25, and 30 months, starting from January 1, 2000, and ending on June 30, 2002.

To make sure we are comparing the KQL version and non-KQL version of the algorithm under the same conditions, we made a few changes to ensure we are only comparing the impact of the KQL layer. For both algorithms, we made sure the query results are stored in a file, instead of directly returning results via memory. We also consolidated the queries in the KQL layer to use a single persistent MongoDB connection for all the SQL queries through the UnityJDBC driver, instead of creating a new connection for each new query. We made this change because the native Python MongoDB driver allows all queries to be executed on the same persistent connection. The KQL version of the algorithm, written in Python, is exactly the same as the non-KQL version, except for replacing the original direct MongoDB query with a KQL query. In the original algorithm, the MongoDB native queries are embedded in the Python calls, and directly passed to the native Python MongoDB driver to directly execute the query. The following experiments were conducted on a MacBook Pro with a 3.06GHz, Intel Core 2 Duo, 8GB of 1067MHz DDR3 RAM, and 512 GB of solid state hard drive.
\begin{figure}
\centering
\includegraphics[width=0.45\textwidth]{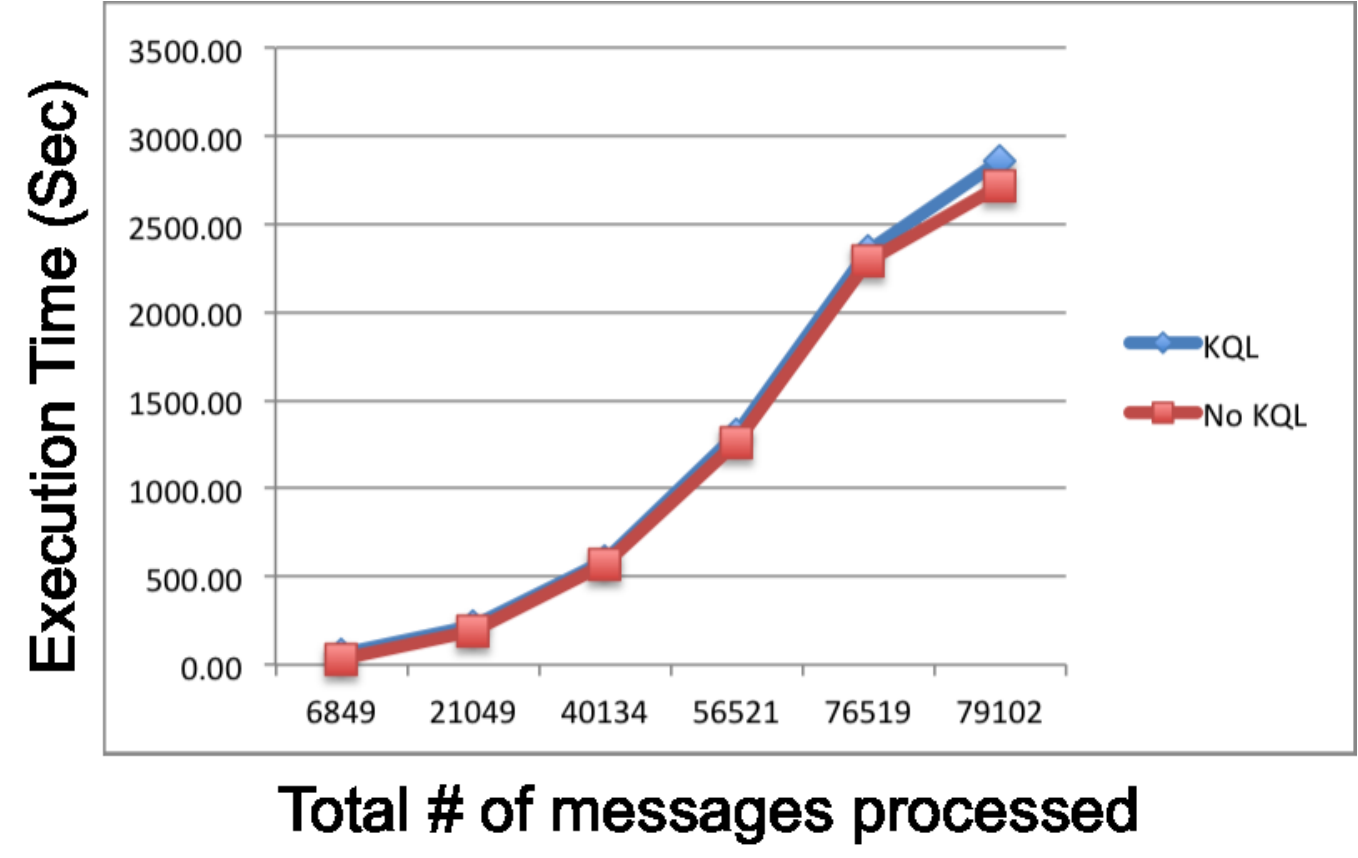}
\caption{Execution time with varying data size.}
\end{figure} 

The first experiment observed execution time while the size of data varied, as the number of messages processed by the algorithm is increased. We chose to increase the number of messages cumulatively over 5 month increments starting from January 1, 2000 till January 31, 2003. The other parameters were set as follows. The minimum number of messages sent by a person is 10, and the number of initial senders is the same as the number of unique sender email addresses in the messages. The comparative execution times are plotted in Fig. 6. The execution time for the NoKQL version ranges from 33 to 2717 seconds, while the corresponding increase in execution time for the KQL version of the algorithm ranges from 18 to 144 seconds under the above settings. Since the smaller data shows higher percentage increase in execution time, we conclude there is some fixed execution time overhead possibly associated with KQL rewriting and Unity JDBC driver, in addition to data size dependent overhead. As the data size increased beyond 40134, the increase in execution time for KQL is under 6\%. The number of KQL calls made in this experiment ranged from 59 to 225 with the average time for a call ranging from 1 sec to 12.7 sec.

For the second experiment, we ran the algorithm to observe the execution time when the initial number of senders varied, while the data size is fixed at 40134 messages, for data collected between January 1, 2000 and March 31, 2001 (see Fig. 7). The execution time with KQL showed varying overhead of up to 25\% with reduced input sizes, with the maximum difference at 26 seconds. The number of KQL calls made in this experiment ranged from 19 to 116 with the average time for a call ranging from 3.5 sec to 5.6 sec.
\begin{figure}
\centering
\includegraphics[width=0.45\textwidth]{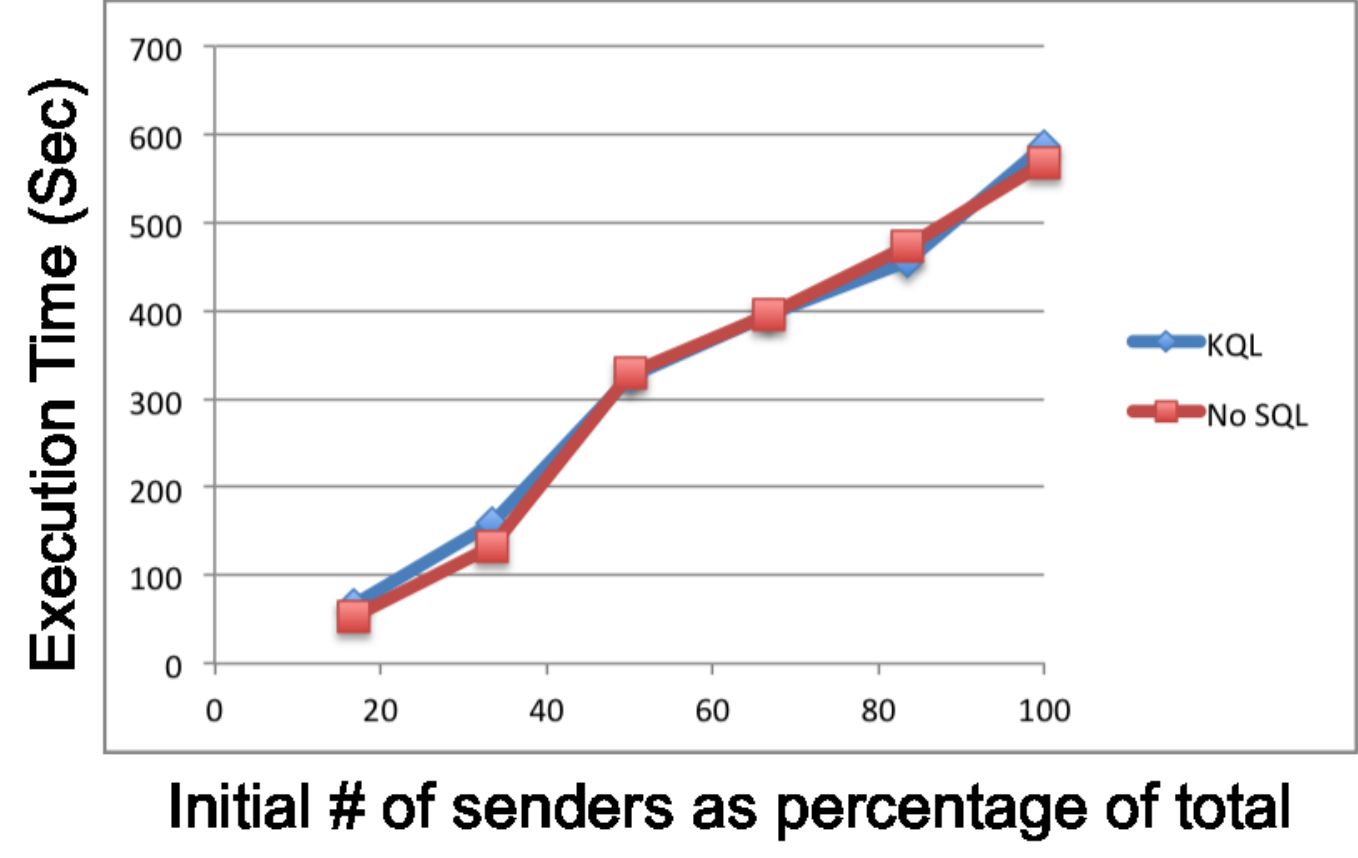}
\caption{Execution time with fixed data size.}
\end{figure} 

The preliminary performance analysis above shows that KQL has overhead that varied depending on the data size and initial input.  More investigation is needed to understand the components of the overhead caused by KQL layer in different circumstances.

\section{Related Work}
Log Analysis is the process of ingesting large raw data logs, storing them efficiently for processing, analyzing the data using queries or algorithms with embedded queries, and finally producing information that an analyst needs. A significant challenge in the log ingestion process is the lack of standardization of log files. 

Even though there have been multiple proposals for log standardization over the years, such as \cite{chuvakin:2008}, the high cost associated with changing the log format is a serious hindrance to industry adopting a logging standard \cite{bitincka2010:optimizing}. Syslog \cite{gerhards:2009rfc} standardizes log content but the format remains text. PCAP is another commonly found format for network packet capture \cite{wir1}. Current commercially available tools for log processing such as Splunk do \cite{bitincka2010:optimizing} little to hide the physical structure and complexity of the underlying logs. 

One approach to querying log files is based on abstractions of the log files. Nascimento et al. \cite{nascimento:2011} show an approach to interpret and find correlations of events by creating an ontology for firewall logs and then using SPARQL \cite{SPARQL:2008} queries on the ontology of logs. However, their technique requires a semantic data store represented in RDF. The works on automated log abstraction, such as \cite{Nagappan:2010, Jiang:2008}, show that the log file record contents can be abstracted into types. Their goal falls short of the creation of a cohesive ontology for logs of a specific domain, such as cyber security.

Another approach to querying logs is using a querying language or process language. Apache Hive \cite{Thusoo:2009} provides a SQL-like language called HiveQL over Hadoop. Apache Pig \cite{PIG} and Apache Mahout \cite{mahout:2011} are two other examples of providing a query interface that is more intuitive to users. Splunk provides a search language based on the UNIX concepts of pipes and commands \cite{bitincka2010:optimizing}. Recent work on extending SQL so that it would work across SQL and NoSQL data stores is also promising \cite{Ong:2014}. In these languages, the query is still carried out without a cohesive ontology that makes porting these search queries portable across installations. 

\section{Conclusions}
In this work, we have demonstrated a feasible alternative approach for standardizing log file content based on its immutable and mutable aspects, using Dimensions, DimensionSets, and Tags, instead of field names of the log records. This approach makes it possible to create queries using SQL with embedded A-Expressions, or Knowledge Query Language (KQL) that are portable across dissimilar physical implementations. Our preliminary performance evaluation shows the overhead caused by the KQL layer is dependent on the data size and the initial input size. Optimizing the implementation of KQL layer is an important future step.

One of the key challenges that any search interface for log files faces is to show the data sources, the tables, and the time period that were used in constructing the query response. These pieces of information, collectively called provenance, need to be returned by the search interface along with the query results. While our current implementation provides limited provenance information, it still needs improvement.  Developing best practices for developing ontology to support KQL is another challenge. Embedding A-Expressions in existing NoSQL languages such as HiveQL \cite{Thusoo:2009} and PIG  \cite{PIG} is also an important future step.
A visual query interface can also be very helpful in building and debugging KQL queries. 

\section{Acknowledgments}
We would like to thank Alexia Schulz for comments and discussions on A-Expressions, and Christine Cunningham for comments. 
This material is based upon work supported by the Assistant Secretary of Defense for Research and Engineering under Air Force Contract No. FA8721-05-C-0002 and/or FA8702-15-D-0001. Any opinions, findings, conclusions or recommendations expressed in this material are those of the author(s) and are not necessarily endorsed by the United States Government.

Delivered to the US Government with Unlimited Rights, as defined in DFARS Part 252.227-7013 or 7014 (Feb 2014). Notwithstanding any copyright notice, U.S. Government rights in this work are defined by DFARS 252.227-7013 or DFARS 252.227-7014 as detailed above. Use of this work other than as specifically authorized by the U.S. Government may violate any copyrights that exist in this work.

\balance

%
\bibliographystyle{abbrv}
\bibliography{references}  
\balancecolumns 
\end{document}